# Chemotaxis and Quorum Sensing inspired Device Interaction supporting Social Networking

Sasitharan Balasubramaniam[1], Dmitri Botvich[1], Tao Gu[2], William Donnelly[1]

Telecommunication Software and Systems Group[1]
Waterford Institute of Technology
Carriganore Campus, Waterford, Ireland
Institute for Infocomm Research[2]
Singapore 119613
{sasib, dbotvich, wdonnelly}@tssg.org[1]
tgu@i2r.a-star.edu.sg[2]

*Abstract*—Conference and social events provides an opportunity for people to interact and develop formal contacts with various groups of individuals. In this paper, we propose an efficient interaction mechanism in a pervasive computing environment that provide recommendation to users of suitable locations within a conference or expo hall to meet and interact with individuals of similar interests. The proposed solution is based on evaluation of context information to deduce each user's interests as well as bio-inspired self-organisation mechanism to direct users towards appropriate locations. Simulation results have also been provided to validate our proposed solution.

## I. INTRODUCTION

Pervasive computing environment ensures transparent computing capabilities that perform decision making on behalf of users with respect to their behaviour and requirements [11] [12]. One application of pervasive service is supporting human interaction by guiding individuals to the right group of people with similar interest during social networking (e.g. conference meetings). The key criteria supporting this application is accurately deducing each user's interests based on context information (e.g. user's interaction history, profile) and physically sorting and clustering individuals of same subject interest. Current techniques for social interaction do not meet these requirements [1] [4]. At the same time current interaction strategy require users to manually search for information of various people to interact with.

Our aim is to create a mechanism to automatically discover people of similar interest and guide individuals towards forming a cluster to support social networking in an environment that consists of large gathering of people. Our approach towards this objective involves two process, which includes (i) evaluation of context information to retrieve Subject and Topic of Interests for each user, and (ii) mechanisms to diffuse this information into the network through P2P signalling to guide users to a specific location and self-organise into physical clusters that represent a particular Subject or Topic of Interest. This paper will concentrate largely on the second mechanism, self-organisation of clusters using bio-inspired techniques.

The paper is organised as follows: Section 2 presents Related Work on current Collaborative Social Interaction techniques as well as examples from Bio-inspired Computing. Section 3 describes key biological principles we have applied towards our solution, while section 4 presents the core components of the architecture as well as algorithms used to self-organise the devices. Section 5 will present preliminary results from simulation work and lastly section 6 will present the Conclusion.

## II. RELATED WORK

### A. Collaborative Pervasive Social Interaction

Gips and Pentland [1] developed a mechanism to allow refinement of user profiles through monitoring their behaviour at conference events. A smart badge is used to monitor the user's behaviour as they migrate through the conference hall. The solution proposed by the authors is to create a dynamic technique to continually refine user's profile, which is not supported by the current mechanisms that require pre-computed static user profiles from data mining of personal data or online forms. Bell et al [4] developed a software architecture that supports recommendation and sharing of software components between users. The system provides recommendation of new system components based on similar history of usage. The architecture has been applied to mobile strategy game where players have the ability to adapt and upgrade their game using components from fellow players. Although the related work described mechanisms to evaluate other user's interest to support interaction, there is no mechanisms that physically guides these remote devices.

### B. Bio-inspired Computing

There are numerous examples of applying cross-disciplinary techniques such as biological principles to computer science. A notable example is the use of biological principles towards robotics [9]. In computer communication Suzuki and Suda [7] have also applied the application of bee colonies for the bio-networking architecture for autonomous applications, where network applications are implemented as a group of autonomous diverse objects called a cyber-entity (CE). Each CE, which is an agent, provides a service to users (e.g., a CE for a web server may contain HTML files), where the CE would reside in a particular node within the network (e.g. router) and within close proximity to the users. The CE exhibits biological behaviours similar to bees such as migration and reproduction. In the event that the service becomes popular, the CE are able to reproduce and migrate towards a node close to the user. The work was later extended to support evolutionary adaptation using genetic algorithms to evolve CE behaviours and improve their survival fitness in the environment [8]. The CE is able to select other CE's with high fitness values and reproduce to form new CEs. The fitness values reflects the popularity of the particular service. As the CE migrates to a node, the CE receives energy from the user using the service, and spends this energy on the node it is residing for utilization of resources. Once the service becomes outdated, the number of users will reduce resulting in the CE spending excessive energy on the node. When the CE's energy is depleted, this results in the termination of the CE.

### III. BIO-INSPIRED MECHANISMS

We apply two Bio-Inspired mechanisms towards our solution, which are chemotaxis as well as quorum sensing.

#### A. Chemotaxis

Chemotaxis is the characteristic movement or orientation of an organism or cell along a chemical concentration gradient either toward or away from the chemical stimulus [10]. Chemotaxis can be both positive or negative, where micro-organism can get attracted to positive chemical gradient towards a particular source (e.g. nutrients) or move away from negative chemical gradient (e.g. moving away from poison).

#### B. Quorum sensing

The process of quorum sensing allows a community of cells to co-ordinate and perform a specific function [2]. An example of quorum sensing is the bacteria *Vibrio fischeri*, which emits light from an organ of a squid, where the cells emit autoinducers to sense the cells within the vicinity. The autoinducers are absorbed by corresponding cells of the same time, and when the autoinducers accumulate beyond a specific threshold, this leads to transcription protein which leads to light production.

### IV. ARCHITECTURE

Our solution is based on evaluating users context profile and deriving a specific *Subject of Interests (*e.g. Pervasive Systems, Computational Intelligence*)* category that the user belongs to. We assume a predefine set of categories for the different Subject of Interests, where each category defines a broad subject. Each Subject of Interests category contains a subset of *Topic of Interests* (e.g. context modeling, smart homes), where the Topic of Interests narrows down the Subject of Interests to a particular field. Our solution for self-organising cluster of devices with similar Subject of Interests and Topic of Interest is based on an architecture illustrated in Fig. 1. The architecture, which is embedded into each device, is composed of two layers, a context management layer as well as a self-organisation layer.

#### A. Context-Management layer

The role of the Context-Management layer is to determine the user's current Subject and Topic of Interest and transfer this information to the self-organising module. This module evaluates and reasons about the user profile which consists of information such as user's occupation, location, as well as user interaction history. In recent years, ontologies have been proven an effective means for modeling user context [3] [5] as they provide rich semantics of the domain knowledge related to a specific area and provide a standard for computers to perform processing and reasoning. We model the user profile using ontology, and adapt the user profile ontology defined in [6] with user interests, preferences, and interaction history. At the same time the history of a user's interaction can be recorded in the user profile to support refinement of the user's interests. For example, if the history of a user's interaction shows that he/she has started interacting with other conference attendees on the topic of smart homes, we may conclude that the user's interest has switched towards this topic.

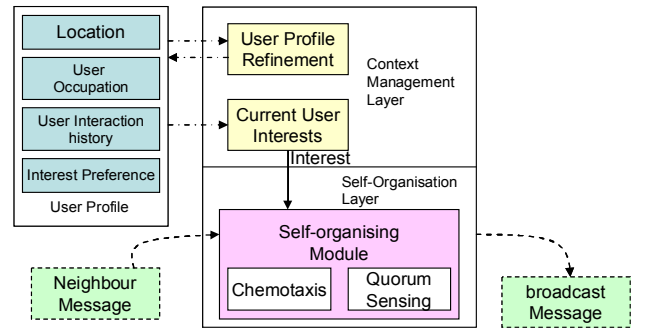

Figure 1. Architecture for device interaction supporting human interests

The matching of users' interests is done by implicitly generating an internal query and performing ontological reasoning [3] [5]. We use Jena APIs [14] to carry out inferences on the ontologies and on local data, and calculate the similarity between the concepts of a query and the concepts of the field terms of annotations defined in the local ontology. The similarity calculation of two sets of concepts is carried out by using the Wu and Palmer formula [13], represented in equation 1,

$$Sim(A,B) = \frac{1}{2}\left(\frac{1}{|A|}\sum_{Ai \in Pl} \max(ConSim(Ai, Bi)) + \frac{1}{|B|}\sum \max(ConSim(Ai, Bi))\right) \quad (1)$$

where $A$ is the set of concepts $\{Ai\}$; $|A|$ cardinal of $A$; $B$: set of concepts $\{Bi\}$; $|B|$ cardinal of $B$; $ConSim(C1, C2))$: similarity calculation function between two concepts $C1$ and $C2$, in a concepts tree, using the equation,

$$ConSim(C1,C2) = 2 \times depth(C)/(depth(C1)+depth(C2)) \quad (2)$$

where, $C$ is the smallest generalization of $C1$ and $C2$ in arcs number and depth $(C)$ is the number of arcs which separates $C$ from the root.

## B. Self-Organisation Layer

The role of the self-organisation layer, illustrated in Fig. 1, is used to self-organise individuals into interest group clusters described previously. The organization of the clusters is based on the principles of chemotaxis and quorum sensing techniques. Fig. 2 illustrates the mechanisms used to self-organise the devices.

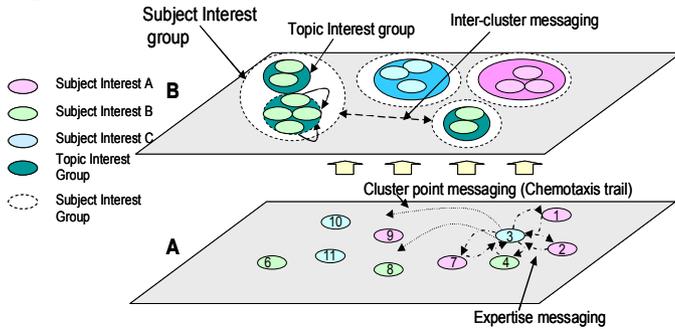

Figure 2. Self-organisation of subject interests

Initially, users enter the conference hall with a mobile device (Fig. 2(A)). Through P2P interaction, the mobile device eventually recommends the user through a Graphical User Interface (GUI) interface the location of the cluster for the Subject and Topic of Interest. As the users move to the recommended point to interact with people with same interest, this leads to clustered groups for each subject interest shown in Fig. 2 (B).

The Self-organisation mechanism, which organises the interests groups into hierarchical clusters, is based on various messaging process that occurs in consecutive stages. The hierarchical cluster structure consists of the primary cluster which represents the Subject of Interests groups, which also contain the sub-clusters that represent specific Topic of Interest. The Algorithm for the Primary cluster formation is illustrated in Fig. 3.

**1 for** $\forall v_i$ **do**
**2**    define $SI$ and calculate $E_{SI}$ for each $v_i$
**3**    **for** $\forall v_i$ broadcast $m_{RE_{SI}}$ **do**
**4**      **if** $\exists v_i \in SI_A : SI_{v_i} \notin SI_{RE_{SI}}$ **then**
**5**        broadcast $m_{RE_{SI}}$
**6**      **else** $\exists v_i \in SI_i : SI_{v_i} \in SI_{RE_{SI}}$
**7**        reply $m_{E_{SI},v_i}$
**8**      **for** $\forall v_i \in SI_i$ receiving $m_{RE_{SI},v_j}$ **then**
**9**        calculate $D_{SI_i,vi}$
**10 for** $\forall v_i$ **do**
**11**    broadcast $m_{D_{SI},v_i}$
**12**    **if** $SI_{v_i} \Leftrightarrow SI_{D_{SI},v_j}$ **then**
**13**      **if** $D_{SI,v_j} < D_{SI,v_i}$ **then**
**14**        broadcast $D_{SI,v_i}$
**15**        maintain pri.-cluster point at $v_i$
**16**      **else**
**17**        $D_{v_{SI_A}} \Leftarrow D_{SI_A,\text{Received}}$
**18**        broadcast $D_{v_{SI_A}}$
**19**        move to $v_j$ pri.-cluster point
**20**    $v_{clusterpoint}$ calculate $D_{SI,v_{clusterpoint}}$
**21**    broadcast $D_{SI,v_{clusterpoint}}$

Figure 3. Algorithm for Self-Organisation of Subject of Interests

The assumption here is that each user posses a single device where each device is categorized into a particular Subjects of Interest $SI$, and there is time synchronization between all the devices to start each process synchronously. Initially, all nodes $v$ within the conference hall will determine the user's $SI$ as well as each user's expertise weight $E_{SI}$ (between 0 – 1.0) (line 2), which indicates how much knowledge and experience each user has with respect to the subject and is evaluated from the Context-Management layer. This is followed by determining the cluster point for each respective subject to set the most appropriate location for the cluster to form. All nodes will initially broadcast an expertise request message $m_{RE_{SI}}$ to the environment. In the event that a node receiving the broadcast message does not belong to the Subject Interest of the broadcasted message, the node will continue to broadcast this message (line 4 - 5). However, if the node does belong to the same Subject of Interest, an *Expertise Message* $m_{E_{SI},v_i}$ is transmitted back. The reply message will also include $E_{SI}$. Once all devices have received the expertise message from nodes with the same Subject of Interest in the surrounding, each device will determine the density weight. This mechanism mimics the quorum sensing process, where the density reflects the concentration of devices with similar subject within the vicinity. The calculation of the density weight is represented by the following equation,

$$D = \alpha \frac{n}{N} + \beta \frac{\sum_{i=1}^{n} EI}{n} \quad (3)$$

The $n$ represents the number of users with the same Subject of Interest from a total of $N$ nodes within the conference Hall. The second term of the equation represents the average expertise weight from the $n$ nodes. Once the density weights are calculated, the cluster point is determined where initially a density message $m_{D_{SI},v_i}$ is broadcast out to the neighbours by each device (line 11). The density message contains the density value and location of the device from where the message originated (we assume here that each device knows their location in the conference hall). Each device evaluates the received density weight with its own weight, and continues to only broadcast the highest density message (lines 12 – 19). This leads to a single direction of broadcast message emitting from the cluster point, resembling a chemotaxis trail. The cluster will slowly form as users with the same Subject of Interest receive the message and move towards the recommended cluster point location. As the cluster increases in size, the chemical concentration on outgoing broadcast messages will increase, where the device that is represented as the cluster point $v_{clusterpoint}$ will continue to calculate and update the new density $D_{SI,v_{clusterpoint}}$. Since multiple clusters with the same Subject of Interest can be formed, Inter-cluster messaging are also transmitted between the clusters to indicate the different topic of interests within the primary clusters. This provide opportunities for users to migrate and join other clusters (line 21). When the users have joined a primary cluster for Subject of Interest, the users begin to interact with participants of the same Subject of Interest. At this point the mechanism to forming the sub-cluster begins. The mechanism to form the sub-cluster follows the same mechanism as the formation of primary cluster. However, the message exchanges are based on Topic of Interests rather than Subject of Interests. Unlike the single Subject of Interests categorization for each user, there can be multiple Topic of Interests for each user. This allows the users to select their most appropriate field and topic to interact within the primary cluster.

## V. SIMULATION

We have evaluated our system through the simulation of a conference scenario. We build our simulation on ns-2 by creating a wireless ad-hoc network with 300 nodes representing 300 attendees for 15 subject interests. Each node is randomly assigned with a weight value (assuming that the density weight has been calculated), a set of interests based on its profile and a location value. The purpose of the simulation is to demonstrate how fast a cluster can be formed based on the chemotaxis trail. As illustrated in Fig. 4, the eventual formation of chemotaxis trail shows a reduction in Cluster Point messaging as the cluster matures with time (the results presented is for a single Subject of Interest). For this single test, we assigned 62 nodes to this particular subject and the time it took to form the cluster was approximately 34 minutes. We have also conducted another simulation experiment to test the effects of cluster formation when the number of nodes per subject of interest varies. The varied number of nodes for each subject interest will determine the time taken to form different subject interest groups with respect to the ratio of the total node population. To simplify the case, we randomly choose one subject for each node, and measure the time taken to form each interest group (note that a user can have multiple subject interests, and might have their own choices to decide which group to join).

Fig. 5 shows that the time taken to form a cluster increases as the number of nodes increase. This is due to the time measured from the first node entering the cluster to the last node entering a cluster of the Subject of Interest. Therefore, in the worst case scenario, for a Subject of Interest of 37 nodes, this will take approximately 18 minutes for the nodes to enter a cluster. While the best case is when there are 5 nodes, which will take approximately 2.8 minutes for each node to enter the cluster.

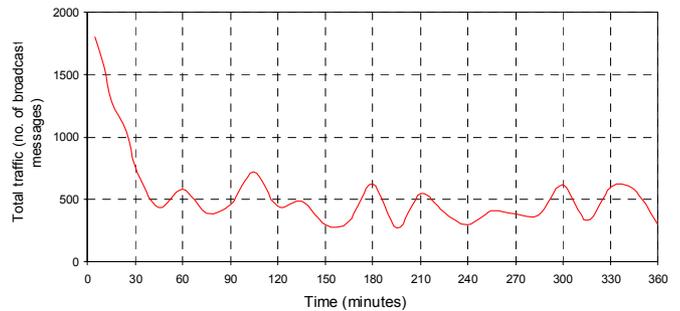

Figure 4. Total network traffic vs. time for forming cluster for single Subject of Interest

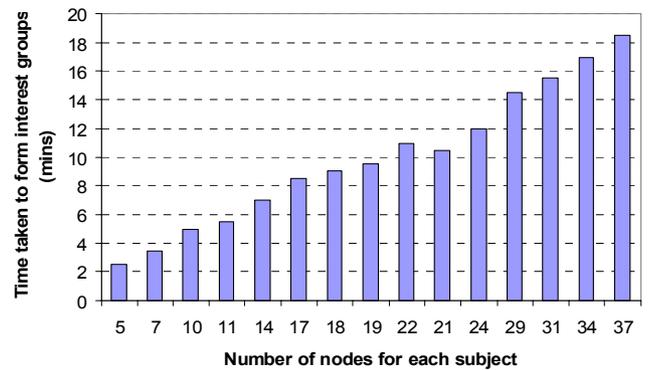

Figure 5. Time taken to form interest groups vs. number of nodes for each subjects

## VI. CONCLUSION

Conference and expo events provide opportunities for people to meet and network with various individuals in order to increase visibilities and create contacts. This paper proposes a pervasive service that efficiently supports users to meet various parties that have similar interests. The mechanism is based on evaluating each user's interest based on context profile and bio-inspired self-organising techniques to enable users of similar interest to automatically come together and form clusters at a specific location in the conference hall. Preliminary simulation results have also been presented to demonstrate the speed of cluster formation for a specific Subject of Interests as well as the speed of cluster formation for varying number of device per Subject of Interest.

REFERENCES

[1] J. Gips, A. Pentland, "Mapping Human Networks," in Proceedings of 5th IEEE Conference on Pervasive Computing and Communications (PerCom 2006), Pisa, Italy, March 2006.

[2] M. B. Miller, B. L. Bassler, "Quorum sensing in Bacteria", Annual Review in MicroBiology, 2001, pp. 165-199.

[3] H. Chen, T. Finin, "An Ontology for a Context Aware Pervasive Computing Environment", in Proceedings of IJCAI workshop on ontologies and distributed systems, Acapulco, Mexico, August 2003.

[4] M. Bell, M. Hall, M. Chalmers, P. Gray, B. Brown, "Domino: Exploring Mobile Collaborative Software Adaptation", in Proceedings of 4th International Conference on Pervasive Computing (Pervasive 2006), Dublin, Ireland, May 2007.

[5] T. Gu, X. H. Wang, H. K. Pung, D. Q. Zhang, "An Ontology-based Context Model in Intelligent Environments", in Proceedings of Communication Networks and Distributed Systems Modeling and Simulation Conference (CNDS '04), San Diego, California, USA, January 2004, pp. 270-275.

[6] M. Golemati, A. Katifori, C. Vassilakis, G. Lepouras, C. Halatsis, "User Profile Ontology version 1", available at http://oceanis.mm.di.uoa.gr/pened/?category=publications.

[7] J. Suzuki, T. Suda, "A Middleware Platform for a Biologically Inspired Network Architecture Supporting Autonomous and Adaptive Applications", IEEE Journal on selected areas in Communications, vol. 23, no. 2, February 2005.

[8] T. Nakano, T. Suda, "Self-organizing Network Services with Evolutionary Adaptation", IEEE Transaction on Neural Networks, vol. 16, no. 5, September 2005.

[9] W. Shen, B. Salemi, P. Will, "Hormone-Inspired Adaptive Communication and Distributed Control for CONRO Self-configurable Robots", IEEE Transaction on Robotics and Automation, vol. 18, no. 5, October 2002.

[10] R. Tyson, L. G. Stern, R. J. LeVeque, "Fractional step methods applied to a chemotaxis model", Journal of Mathematical Biology, vol. 41, 2000, pp. 455 – 475.

[11] M. Satyanarayanan, "Pervasive Computing: Vision and Challenges", IEEE Personal Communication, vol. 8, no. 4, August 2001, pp. 10 – 17.

[12] M. Weiser, "The Computer for the Twenty-First Century", Scientific American, 265, September 1991, pp. 94-104.

[13] Z. Wu, M. Palmer, "Verb Semantics and Lexical Selection". In Proceedings of 32nd Annual Meeting of the Association for Computational Linguistics, 1994, pp. 133–138.

[14] Jena 2 – "A Semantic Web Framework", http://www.hpl.hp.com/semweb/jena2.htm.